\documentclass[12pt]{article}
\usepackage{cite}

\makeatletter
\@addtoreset{equation}{section}

\makeatletter
\renewcommand\section{\@startsection {section}{1}{\z@}%
                                   {-3.5ex \@plus -1ex \@minus -.2ex}
                                   {2.3ex \@plus.2ex}%
                                   {\normalfont\large\bfseries}}
\renewcommand\subsection{\@startsection{subsection}{2}{\z@}%
                                     {-3.25ex\@plus -1ex \@minus -.2ex}%
                                     {1.5ex \@plus .2ex}%
                                     {\normalfont\bfseries}}

\def\baselinestretch{1.2}
\newcommand{\be}{\begin{equation}}
\newcommand{\ee}{\end{equation}}
\newcommand{\beq}{\begin{eqnarray}}
\newcommand{\eeq}{\end{eqnarray}}

\begin{document}
\begin{titlepage}
\begin{flushright}
hep-th/0401197\\
MAD-TH-04-1\\
MCTP-04-04\\
\end{flushright}

\vfil\

\begin{center}

{\large{\bf Correspondence Principle for Black Holes in Plane Waves}}
\vfil

\vspace{3mm}

Akikazu Hashimoto$^a$ and Leopoldo Pando Zayas$^b$

\vspace{8mm}

$^a$ Department of Physics\\
University of Wisconsin, Madison, WI 53706\\

\vspace{8mm}

$^b$ Michigan Center for Theoretical Physics \\
Randall Laboratory of Physics\\
University of Michigan, Ann Arbor, MI 48109-1120

\vfil

\end{center}

\begin{abstract}
\noindent We compare the entropy as a function of energy of excited
strings and black strings in an asymptotically plane wave background
at the level of the correspondence principle.  For the plane wave
supported by the NSNS 3-form flux, neither the entropy formula nor the
cross-over scale is affected by the presence of the flux and the
correspondence is found to hold. For the plane wave supported by the
RR 3-form flux, both the entropy and the cross-over point are
modified, but the correspondence is still found to hold.
\end{abstract}
\vspace{0.5in}

\end{titlepage}
\renewcommand{\baselinestretch}{1.05}  

\section{Introduction}

Plane waves have received increased attention recently as an important
background space-time in string theory.  These backgrounds are
interesting for two reasons. On one hand, the theory on the world
sheet admits a simple realization, making many explicit computations
possible \cite{Russo:1995cv} even in the presence of a Ramond-Ramond
background flux \cite{Metsaev:2001bj}. On the other hand, certain
plane wave backgrounds \cite{Blau:2001ne} admit dual interpretation as
a scaling limit of certain field theories
\cite{Berenstein:2002jq}. This provides an exciting opportunity to
explore gravitational physics, in an asymptotically plane wave
background geometry, strictly in the framework of quantum field
theory.

The most immediate application of this duality in addressing problems
of gravity that comes to one's mind is the physics of Schwarzschild
black holes.  Consideration of Schwarzschild black holes in an
asymptotically anti de-Sitter space-time played a significant role in
clarifying the nature of holography in the context of AdS/CFT
correspondence \cite{Witten:1998zw}, and it is natural to expect that
Schwarzschild black holes in an asymptotically plane wave space-time
would equally clarify the nature of the correspondence of
\cite{Berenstein:2002jq}.  The first step in such a line of
investigation is to construct an explicit black hole solution.
Immediately following the proposal of \cite{Berenstein:2002jq}, there
have been numerous attempts to construct solutions of this type, with
precisely this goal in mind.

Constructing Schwarzschild black hole solutions in an asymptotically
plane wave space-time has proven to be extremely challenging.  The
difficulty stems from the fact that the symmetries of a background
with a horizon and an asymptotically null background fluxes are not
transparent. A review of the structure of horizons and plane waves can
be found in \cite{Hubeny:2003ug}.

The first explicit solution of Schwarzschild black strings in an
asymptotically plane wave solution was written down very recently in a
series of papers \cite{Gimon:2003ms,Gimon:2003xk}. (An explicit
solution of a BPS black string in an asymptotically plane wave
geometry was identified earlier in an inspiring paper
\cite{Herdeiro:2002ft}.)  Unfortunately, these black strings are
embedded in a wrong asymptotic plane wave for the correspondence of
\cite{Berenstein:2002jq} to be applicable.  This correspondence
relates a limit of ${\cal N}=4$ SYM to type IIB string theory on a
plane wave supported by the Ramond-Ramond {\em five-form flux}
\cite{Blau:2001ne}. The solutions constructed in
\cite{Gimon:2003ms,Gimon:2003xk} are that of black strings embedded in
a plane wave supported either by the NSNS or the RR {\em three-form
fluxes} \cite{Bena:2002kq,Michelson:2002ps}. We are therefore unable
to explore the properties of these black holes by studying the ${\cal
N}=4$ SYM and using the correspondence of \cite{Berenstein:2002jq}.

Despite this shortcoming, plane waves supported by three-form fluxes
are still special in that the world sheet theory is simple. Of course,
the application of full interacting string theory as a means to study
the microscopic properties of Schwarzschild black holes is an
important outstanding challenge, even if the world sheet theory is
extremely simple (such as in the case of Minkowski space).  There
does exist, however, a scheme to semi-quantitatively compare the
properties of {\em free} strings and black holes, sometimes referred
to as the {\em correspondence principle}
\cite{Susskind:1993ws,Horowitz:1997nw}. The goal of this article is to
explore the black string solutions of \cite{Gimon:2003ms,Gimon:2003xk}
in this context.

Correspondence principle in plane wave space-time was also considered
in \cite{Li:2002mq}, but its content appears to have very little
overlap with what is presented in this paper.

The organization of this paper is as follows. We begin in section 2 by
reviewing the construction of black string solutions
\cite{Gimon:2003ms,Gimon:2003xk}.  In section 3, we review the
correspondence principle and how it applies to black strings in plane
waves supported by the NSNS 3-form field strength. In section 4, we
describe the string theory computation relevant for the application of
the correspondence principle. In section 5, we examine the status of
the correspondence principle for the plane waves supported by the RR
3-form field strengths. We will conclude in section 6.

It should be emphasized that the construction of a black hole/black
string solution in an asymptotically plane wave space-time supported
by the RR five-form remains an important outstanding problem.  We hope
that this problem will be solved in due course and will pave a path
toward a quantitative study of black hole physics in terms of the
${\cal N}=4$ SYM theory.

\section{Black Strings in Asymptotically Plane Wave Space-Time}

In this section, we will review the construction of black string
solutions first reported in \cite{Gimon:2003ms,Gimon:2003xk}.  The
main ingredient behind this construction is a set of manipulations
which was originally formulated in \cite{Alishahiha:2003ru} which we
refer to as {\em Null Melvin Twist} following \cite{Gimon:2003xk}. The
reader should refer to \cite{Gimon:2003xk} for the details of this
manipulation.  Here, we summarize the steps using a slightly different
notation.
\begin{enumerate}
\item Consider a ten dimensional Minkowski space-time for a type IIB
supergravity theory, written in terms of coordinates $t$, $y$,
$\rho_i$, $\phi_i$:
\be ds^2 = -dt^2 + dy^2 + \sum_{i = 1}^4 (d \rho^2 + \rho^2 d
\phi_i^2)\ee

\item Boost to a new frame
\be \left( \begin{array}{c}t \\ y \end{array} \right)
= \left( \begin{array}{cc} \cosh \gamma & -\sinh \gamma \\ -\sinh \gamma & \cosh \gamma \end{array}\right) \left( \begin{array}{c}t' \\ y' \end{array} \right)
\ee
\item Compactify $y'$ so that it has radius $R$ and T-dualize along
$y'$ so that the new coordinate $\tilde y'$ has radius $\alpha'/R$.
\item ``Twist,'' by replacing $d\phi_i $ by $d \phi_i + \omega_i\, d
\tilde y'$ in the line elements
\item T-dualize from $\tilde y'$ to $y'$

\item Boost back to the original frame
\be \left( \begin{array}{c}t' \\ y' \end{array} \right)
= \left( \begin{array}{cc} \cosh \gamma & \sinh \gamma \\ \sinh \gamma & \cosh \gamma \end{array}\right) \left( \begin{array}{c}t \\ y \end{array} \right)
\ee
\end{enumerate}
At this stage, one arrives at a seemingly complicated space-time
depending on parameters $R$, $\gamma$ and $\omega_i$ for $i=1\ldots4$.
All of these space-times, however, are related to Minkowski space by
dualities, boosts, and twists, and belong to the class of exactly
solvable string theories considered \cite{Russo:1995tj,Russo:1996ik}.
For example, setting $\gamma=0$ gives rise to the general NSNS Melvin
solution.  If instead we set all $\omega_i = \omega$ and
\begin{enumerate}
\item[7.] Scale $\gamma$ to infinity, keeping
\be \eta = {1 \over 2} \omega e^\gamma = \mbox{fixed} \ . \ee
\end{enumerate}
Then, the end result is a plane wave geometry
\beq ds^2 & = & -dt^2 + dy^2  - \eta^2 r^2 (dt + dy)^2 + \sum_{i=1}^4  (d \rho_i^2 + \rho_i^2 d \phi_i^2) \cr
e^\varphi & = & 1 \label{planewave}\\
B & = & \eta (dt + dy) \wedge (\sum_{i=1}^4 \rho_i^2 d \phi_i) \nonumber
\eeq
where $r^2 = \sum_i \rho_i^2$.

Although we have so far only described Null Melvin Twists applied to
Minkowski space-time, these manipulations can easily be applied to any
space-time with a translational isometry for T-dualizing and a
rotational isometry for twisting. One can, for example, apply the Null
Melvin Twist to the Schwarzschild black string solution
\be ds^2 = -f(r) dt^2 + dy^2 + {1 \over f(r)} dr^2 + r^2 d \Omega_7^2 \label{blackstring}\ee
\be f(r) = {1 - {M \over r^6}} \ .\ee
To apply the Null Melvin Twist, it may be more convenient to rewrite
\be {1 \over f(r)} dr^2 + r^2 d \Omega_7^2 \qquad \mbox{as}  \qquad  {1 - f(r)\over f(r)} dr^2 + \sum_i (d \rho_i^2 + \rho_i^2 d \phi_i^2) \ . \ee
In the end, one obtains the space-time
\beq
ds_{str}^2 & =& - {f(r)\, \left(1 + \eta^2 \, r^2
\right) \over k(r)} \, dt^2 -  \, {2 \, \eta^2 \, r^2 \, f(r) \over k(r)} \,
dt \, dy + \left( 1  -{ \eta^2\,  r^2 \over k(r)} \right) \, dy^2  \cr
& & \qquad  + {dr^2 \over f(r)} + r^2 \, d\Omega_7^2 - {\eta^2 \, r^4 
\, (1 - f(r))
\over 4 \, k(r)}\, \sigma^2  \label{ppblackstring}\\
e^{\varphi} &=& { 1 \over \sqrt{k(r)}} \cr
B & = &{\eta \, r^2 \over 2 k(r)}\, \left(f(r) \, dt + dy\right) \wedge
\sigma  \nonumber \eeq
where
\be k(r) = 1 + {\eta^2 M  \over r^4} \ , \qquad \mbox{and} \qquad r^2 \sigma = {1 \over 2} \sum_{i=1}^4  \rho_i^2 d \phi_i \ . \ee
This is the supergravity solution for a Schwarzschild black string in
an asymptotically plane wave space-time supported by an NSNS three
form flux, whose magnitude is parameterized by $\eta$.  A similar
solution supported by the RR three from flux can be constructed
immediately by S-dualizing the supergravity solution. These are the
solutions that we will be considering in this paper.

Let us comment about the physical properties of these black strings.
Among the most basic physical characteristics of black strings are its
horizon area and its surface gravity. The horizon for the space-time
(\ref{ppblackstring}) is located at $r_H^6 = M$. Its area in Einstein
frame is readily computed to be
\be {\cal A} = 2 \pi R\,   M^{7 / 6} \Omega_7 \ , \ee
where we have assumed that the $y$ coordinate is
compactified\footnote{Such a compactification gives rise to closed
time-like curves and may lead to additional subtleties of the type
considered in \cite{Brace:2003st}. We will not be addressing this
point in this paper.} on a circle of radius $R$, and $\Omega_7$ is the
area of the unit 7-sphere. This area is independent of the background
field strength parameterized by $\eta$.

Computing the surface gravity is a bit more subtle. It is defined as
\be \kappa^2 = -{1 \over 2} (\nabla^a \xi^b) (\nabla_a \xi_b) \ee
in Einstein frame, where $\xi^a$ is the time-like Killing vector
normal to the horizon (See e.g.\ \cite{Wald:1984rg}). In the case of
solution (\ref{ppblackstring}), the vector
\be \xi^a = \left( {\partial \over \partial t} \right)^a \label{killing} \ee
turns out to be the one normal to the horizon.  For this choice of the time-like Killing vector, the surface gravity 
\be \kappa^2 = {36 \over M^{1/3}} \ee
also comes out to be independent of the parameter $\eta$.  Of course,
the precise value of the surface gravity and temperature depends on
the normalization of the Killing vector $\xi^a$.  We will {\em choose}
to normalize $\xi^a$ precisely as (\ref{killing}) in the coordinate
where the metric asymptote to a geometry of the form given in
(\ref{ppblackstring}). This means that in order to make a sensible
comparison between black hole thermodynamics and the statistical
mechanics of a microscopic system, one must evaluate the Boltzmann
trace of the specific Hamiltonian operator conjugate to the Killing
vector (\ref{killing}).

\section{Review of the Correspondence Principle}

The correspondence principle states that in a weakly coupled string
theory, an object of mass $m$ has the physical properties of a
classical black hole to a good approximation as long as the curvature
of the metric in string frame near the horizon is smaller than the
scale set by the string scale, and has the physical properties of an
excited string if the Schwarzschild radius is smaller
\cite{Susskind:1993ws}. As a consequence of this principle, at the
critical mass $m$ where the Schwarzschild radius is of the order of
the string scale, the entropy of the excited string in a flat
background and the entropy of the black hole must also be of same
order of magnitude \cite{Horowitz:1997nw}.

This principle can easily be verified for black holes and black
strings in asymptotically flat space-time. Let us consider, for the
sake of concreteness, a neutral black string in ten space-time
dimensions (\ref{blackstring}) wrapping a compact circle of radius
$R$, which is equivalent to a black hole in nine space-time
dimensions.  The square of the Riemann curvature tensor $R_{\mu \nu
\lambda \sigma} R^{\mu \nu \lambda \sigma}$ scales as $1 / r_H^4$,
indicating that the expected cross-over scale is when $r_H$ is of the
order of $l_s$.\footnote{To be more precise, the square of Riemann
tensor does have a mild dependence on $\eta$, behaving like
$f(\eta)/r_H^4$ where $f(\eta)$ is a function
which smoothly interpolates between numbers of order one as $\eta$ is
varied from zero to infinity.}

The entropy of such a black hole is easily computed to be
\be S_{BH} = {1 \over G_9} (G_9 E)^{7/6} \ee
where $E$ is the energy associated with the black string 
\be E =  {M \over G_9}  \ee
and is derived using the standard relation  $d E = T d S$.
$G_9$ is the nine dimensional Newton constant
\be G_9 = {G_{10} \over  R}  = {g_s^2 l_s^8  \over R} \ . \ee
The entropy of a string is dictated by the Hagedorn density of states
\be S_s = l_s E \ . \label{stringentropy} \ee
We have ignored the numerical factor of order one in this estimate as
they are irrelevant in the context of the correspondence principle.
The Schwarzschild radius is
\be r_H^6 = G_9 E \ . \ee
One can immediately verify that the entropies
\be S_{BH} = S_{s} = {l_s^7 \over G_9} \ee
match when 
\be l_s^6 = r_H^6  = G_9 E \ . \ee
This relation holds for any value of $g_s$ provided that its value is
of order one or less. This is the essence of the statement of the
correspondence principle for black strings in an asymptotically flat
space-time.

One additional technical comment is in order regarding the
Gregory-Laflamme instability of the black string solution.  Black
strings are unstable to decay when the Schwarzschild radius $r_H$ is
of the order of the compactification radius $R$. This means that the
black string is unstable at energies below $E = R^6/G_9$.  In order
for this instability to be hidden below the cross-over scale, $R$ must
be of the order or smaller than $l_s$. Of course, if $R$ is smaller
than $l_s$, one must worry about the Gregory-Laflamme instability of
the T-dual picture. We are therefore forced to consider the case where
the compactification radius $R$ is of the order of the string scale
$l_s$ in order to apply the correspondence principle to the black
string solution.

In the case of the black strings in an asymptotically plane wave
background (\ref{ppblackstring}), we found in the previous section
that the area of the horizon and the surface gravity are unaffected by
the three form field strength parameter $\eta$.  This implies that the
entropy and the temperature are also unaffected by $\eta$.  In order
for the critical mass at the cross-over point to correspond to a black
string whose Schwarzschild radius is the string scale, the entropy of
the strings in the plane wave background should also be unaffected by
$\eta$. This is what we will examine in the following section.

\section{Thermal Partition Functions for Strings in Plane Wave Geometry}

In this section we will describe the computation of the thermal
partition function of strings in asymptotically plane wave background
from which one can derive the formula for the entropy which enters in
the consideration of the correspondence principle.  Our goal is to
compute the Boltzmann trace over for the spectrum of generator
(\ref{killing}) for the set of string states in these backgrounds.
Our task is drastically simplified in light of the fact that
quantization of the strings and the computation of closely related one
loop partition functions have been done already by many people
\cite{Russo:1995cv,Russo:1995tj,Russo:1996ik,Takayanagi:2001jj,Dudas:2001ux,Takayanagi:2001aj,Takayanagi:2002je,Takayanagi:2002pi,PandoZayas:2002hh,Greene:2002cd,Sugawara:2002rs,Brower:2002zx,Sugawara:2003qc,Grignani:2003cs,Hyun:2003ks,Bigazzi:2003jk}. For
the purpose of illustration, let us work with the bosonic theory and
follow the notation of \cite{Russo:1995tj}. This paper considered
string theory in a background defined by the sigma-model of the form
\be I = {1 \over \pi \alpha' \tau_2} \int d^2 \sigma\, [ F^{-1}(x) C \bar C  + \bar C ( \partial u' + A_1 ) - C (\bar \partial v' + A_2) + \sum_i \partial x'_i \bar \partial x'^*_i ]  \label{background} \ee
where
\beq A_1 &=& \partial y_* - {i \over 2} \sum_i  \alpha_i (x'_i \partial x'^*_i - x'^*_i \partial x'_i) \cr
A_2 & = & \bar \partial y_* + {i \over 2} \sum_i  \beta_i (x'_i \bar\partial x'^*_i - x'^*_i \bar\partial x'_i) \cr
F^{-1}(x) & = & 1 + \sum_i \alpha_i \beta_i x'_i x'^*_i \cr
x'_i & = & e^{i (q_{+i} y + q_{-i} t)} x_i \ .
\eeq
We have slightly generalized the model of \cite{Russo:1995tj} to
allow non-vanishing values of $\alpha_i$, $\beta_i$, $q_{+i}$, and
$q_{-i}$ for each of the 12 transverse planes of the bosonic string
theory.  Setting
\be \alpha_i = 2 \eta, \qquad \beta_i = 0, \qquad  q_{+i} = \eta,\qquad q_{-i} = -\eta \ .  \label{parameters} \ee
gives rise to precisely the 26 dimensional version of the plane wave
geometry (\ref{planewave}).

The vacuum partition function in this background is 
\be Z_0 = \int_{{\cal F}} {d^2 \tau \over \tau_2} \sum_{m,w} \int {d \varepsilon \over 2 \pi}  {\rm Tr} \left( e^{2 \pi i (\tau L_0 - \bar \tau \bar L_0)} \right) \ee
where 
\be  \varepsilon = -i E \ee
is the imaginary extension of the zero mode along the time coordinate
and sum over $m$ and $w$ corresponds to the momentum and the winding
number along the compact $y$ coordinates. The $\tau$ integral is done
over the fundamental domain ${\cal F}$.

To compute the thermal partition function, on the other hand, we must
evaluate
\be Z_T = \left. {1 \over T} \int_{{\cal E}} {d^2 \tau \over \tau_2} \sum_{m,w} \int {d \varepsilon \over 2 \pi} e^{i k' \varepsilon / T} {\rm Tr} \left( e^{2 \pi i (\tau L_0 - \bar \tau \bar L_0)} \right) \right|_{k'=-1} \label{thermopartition1}\ee
where ${\cal E}$ refers to the semi-infinite strip $0 < \tau_1 < 1$,
$0 < \tau_2 < \infty$.  To see that this is the thermal partition
function, write
\be 2 \pi i (\tau L_0 - \bar \tau \bar L_0) 
= - 2 \pi \tau_2 (L_0 + \bar L_0) + 2 \pi i \tau_1 (L_0 - \bar L_0) \ee
where the term proportional to $\tau_2$ can be rewritten
\be L_0 + \bar L_0 = {\alpha' \over 2} (-E^2 + A E + B) \ee
where $A$ and $B$ are simultaneously diagonalizable operator acting on
the space of oscillating strings. Integrating out $\varepsilon$,
$\tau_1$, and $\tau_2$ gives rise to
\be Z_T =  {\rm Tr} \left( \exp\left( - {1 \over T} \left(\sqrt{B + {A^2 \over 4}} + {A \over 2}\right)\right)  \delta_{L_0 - \bar L_0} \right) \ee
which can readily be interpreted as the Boltzmann sum over the
ensemble of excited single string states. This form of writing the
thermal partition function can be related to the formulation in terms
of light-cone Hamiltonian as it appears for example in
\cite{PandoZayas:2002hh} by starting from (\ref{thermopartition1}) and
following the sequence of steps
\begin{enumerate}
\item Decompactify the $y$ coordinate by replacing $m$ by $R\, p_y$,
setting $w=0$, and sending $R$ to zero.
\item Change variables by setting $p_y = p_+ - E$
\item Set $E = i \varepsilon$ and integrate out $\varepsilon$ and $\tau_2$.
\end{enumerate}
Decompactification however will give rise to a diverging volume
factor. While overall normalization of the partition function is not
particularly important for computing the density of states, we will
make the point of compactifying $y$ the coordinate on finite spatial
circle, at least near $r=0$.

Details of the computation of the thermal partition function will be
presented in the appendix. Here, let us simply quote the result that
for the general model, the thermal partition function takes the form
\beq Z_T &=& {R V_{22} \over (2 \pi)^{23}\alpha'^{12} T}  \int_{{\cal F}} {d^2 \tau \over \tau^{2}_{2}}   \, 
  \sum_{w,w'=\-\infty}^\infty  \left( 4 \tau_2^{-1} \int d \lambda d \bar \lambda \right)  
 \left( {e^{4 \pi \tau_2} \over \tau_2^{12}}    |f(e^{2 \pi i \tau})|^{-48} \right) \cr
&& \prod_{i=1}^{12} \left({1 \over (2 \pi)^2} \exp[-{\pi(\chi_i - \tilde \chi_i)^2 \over 2  \tau_2} ] {{\tau_2} |\theta'_1(0,\tau)|^2 \over \theta_1(\chi_i| \tau) \theta_1(\tilde \chi_i| \bar \tau)}  \right) \cr
&& \exp\left( - 4 \pi \tau_2^{-1} \left(\lambda \bar \lambda 
- \left({1\over 2} r (w' - \tau w)+{i (k' - \tau k) \over 4 \pi \sqrt{\alpha'} T} \right) \bar \lambda  \right. \right. \label{final3}\\
&& \left. \left. \qquad
+ \left({1\over 2} r (w' - \bar \tau w) - {i(k' - \bar \tau k) \over 4 \pi \sqrt{\alpha'} T}\right)\lambda
\right) \right) \nonumber    \eeq
where
\beq
\chi_i & = & - \sqrt{\alpha'} [2 \beta_i \lambda  + q_{+i} r(w' - \tau w)] + {i q_{-i} (k' - \tau k)\over 2 \pi T} \cr
\tilde \chi_i & = &  -\sqrt{\alpha'} [2 \alpha_i  \bar \lambda  + q_{+i} r(w' - \bar \tau w)]  + {i q_{-i} (k' - \bar \tau k) \over 2 \pi T} \ .  
\eeq

The Hagedorn temperature associated with this thermal partition
function can be extracted by studying the large $\tau_2$ asymptotics
of the integrand of the $\tau$ integral \cite{Atick:1988si}.  In this
limit,
\be
f(e^{2 \pi i \tau})  \rightarrow  1, \qquad 
{\theta_1'(0,\tau) \over \theta_1(\chi_i,\tau)}  \rightarrow  {\pi \over \sin(\pi \chi_i) }, \qquad 
\exp[-{\pi(\chi_i-\tilde \chi_i)^2 \over 2 \tau_2}]  \rightarrow  1\ , 
\ee
so that the $\lambda$ integral becomes Gaussian. After doing this
integral, The Hagedorn temperature can be extracted by comparing the
growth of $e^{4 \pi \tau_2}$ and the decay of
\be \exp\left(-4 \pi \tau_2^{-1} {(\tau k) (\bar \tau k) \over 16 \pi^2 \alpha' T^2} \right)  \sim \exp\left(- \tau_2 {k^2 \over 4 \pi \alpha' T^2}\right) \ .\ee
The critical temperature $T_{crit}$  comes out to 
\be T_{crit} = {k \over 4 \pi \sqrt{\alpha'}} \ , \ee
which takes the  smallest value 
\be T_H = {1 \over 4 \pi \sqrt{\alpha'}} \ee
for $k=1$. This result is independent of $\eta$ which only enters the
partition function through the values of $\alpha_i$, $\beta_i$,
$q_{+i}$, and $q_{-i}$. So we find that the Hagedorn temperature for
the plane wave is the same as the Hagedorn temperature of bosonic
strings in Minkowski space. This was in fact pointed out first for the
case of backgrounds compactified along a light-like direction in
\cite{Sugawara:2002rs}. We therefore conclude that the entropy of
strings in (\ref{planewave}) is also given by
(\ref{stringentropy}). Since neither the black string entropy nor the
string entropy were modified by the presence of $\eta$, the two will
be of the same order of magnitude precisely when the Schwarzschild
radius is of the order of the string length. In short, the
correspondence principle is working.

\section{Correspondence Principle for Black Strings in a Plane Wave with Ramond-Ramond Flux}

So far, we have considered the cross-over in the physical properties
of excited strings and a black string in the background of plane wave
supported by the NSNS three form flux. Let us now consider the same
issue for the case of plane wave background supported by the RR three
form flux.

On the black string side, the supergravity solution corresponding to a
black string in an asymptotically plane wave background with RR three
form flux with strength $\mu$ can be constructed straightforwardly by
S-dualizing the solution (\ref{ppblackstring}).  As far as the
physical properties such as the horizon area and the surface gravity
is concerned, one should work with these space-time in the Einstein
frame.  The Einstein metric, however is invariant under the S-duality
transformation, leading to the conclusion that the entropy and the
temperature is independent of the strength of the background
Ramond-Ramond three form $\mu$.

On the perturbative string side, the generalization of (\ref{final3})
for type IIB theory with RR-background can also be computed and takes
the form
\be
Z_T  =   {R \over (2 \pi)^2  \alpha' T}\int_{{\cal F}} {d^2 \tau \over \tau_2^2}\,  \sum_{\epsilon_i = 0,1} \sum_{{k \in 2{\bf Z} + \epsilon_1 \atop  k' \in 2{\bf Z}+ \epsilon_2}} \sum_{w,w' \in {\bf Z}} e^{ - {1 \over 4 \pi \alpha' \tau_2 T^2} - {\pi R^2\over \tau_2 \alpha'}(w' + \tau w) (w' + \bar \tau w)} \,  
Z^{tr}_{\epsilon_1,\epsilon_2}(\tau,\bar \tau ; m) \ , 
\ee
where\footnote{Note that we are using $\mu$ which differ in
normalization with what was used in \cite{Sugawara:2002rs} by a factor
of $\sqrt{2}$.}
\beq 
m^2 &=& \mu^2 \left({1 \over (2 \pi)^2 \tau_2^2 T^2} (k' - \tau k)(k' - \bar \tau k) -
{i  R \over 2 \pi \tau_2^2 T} \left(2 k w \tau \bar \tau - (\tau+\bar\tau)(k' w + k w') + 2 k' w'\right) \right. \cr
&&\qquad\qquad  \left. - {R^2 \over  \tau_2^2} (w' - \tau w)(w' - \bar \tau w) \right) \ ,
\eeq
and
\be Z^{tr}_{a,b}(\tau,\bar \tau ; m) = {\Theta_{(a,b)}(\tau , \bar \tau; m) ^4 \over \Theta_{(0,0)}(\tau,\bar \tau; m)^4} \ , \ee
as was defined in \cite{Takayanagi:2002pi,Sugawara:2002rs}.  This
partition function corresponds to space-like compactification of the
$y$ coordinate near $r=0$ so as to provide a natural normalization to
the partition function. It is related to the light-cone compactified
case of \cite{Sugawara:2002rs} by infinite boost accompanied by a
scaling of $R$ and $\mu$.  To take the flat space limit, one should
replace the divergent factor
\be Z_0 = {1 \over (2 \pi)^2  \tau_2^2 m^2} = {1 \over (2 \pi)^2 \tau_2^2 m^2} \ , \ee
which comes from the trace over the bosonic zero-mode oscillator
\be Z_T = Z_0 Z'_T \ , \ee
by the standard zero-mode factor
\be Z_0 = {V_2 \over (2 \pi)^2 \tau_2 \alpha'} \ . \ee

The Hagedorn temperature can be computed by analyzing the large
$\tau_2$ behavior of the $\tau$ integral as before. This time,
however, one finds that the Hagedorn temperature, given implicitly by
the relation
\be {1 \over 8 \pi^2 \alpha' T_H^2} - 8 \left( \Delta\left({\mu \over 2 \pi T_H} ; {1 \over 2} \right) - \Delta \left( {\mu \over 2 \pi T_H }; 0 \right) \right) = 0 \ ,  
\ee
{\em does} depend on the strength of the background RR flux $\mu$, as
was found earlier in similar models
\cite{PandoZayas:2002hh,Greene:2002cd,Sugawara:2002rs,Brower:2002zx}. We
have used the notation of \cite{Sugawara:2002rs} where
\be \Delta(m;a) = -{1 \over 2 \pi^2} \sum_{n=1}^\infty \int_0^\infty ds \, e^{-s n^2 - {\pi^2 m^2 \over s}} \cos(2 \pi n a) \ . \ee
For small $\mu l_s$, the Hagedorn temperature $T_H$ is of the order of
the string scale $1/l_s$.  For large $\mu l_s$, $T_H$ grows like
\be T_H(\mu) \sim {\mu \over 2 \log (\mu l_s)} \left(1 + {\cal O}(1/\log(\mu l_s)) \right) \ . \ee

To test the correspondence principle, first note that the entropy of
the black hole
\be S = {1 \over G_9} (G_9 E)^{7/6} \ee
and the entropy of the excited strings
\be S = {E \over T_H(\mu)} \ee
are in agreement precisely when
\be r_H = (G_9 E)^{1/6} = {1 \over T_H(\mu)} \ . \ee
The correspondence principle requires that this is precisely the point
where the curvature of the black-hole solution gets large in the unit
set by the dynamics which gives rise to fluctuation in the background.
In a sufficiently weakly coupled string theory, this fluctuation is
associated with the stringy halo.  The scale of this halo is set by
the scale of the exponential growth in the density of stringy
excitations. So if the correspondence principle is working, one
expects the curvature near the horizon of the black hole to be of the
order of magnitude set by $T_H(\mu)$ when the radius of the horizon is
of the same order $r_H = 1/T_H(\mu)$.

By explicit calculation, one finds that the curvature of the black
string solution in the Ramond-Ramond plane wave background scales like
\be R^2(r_H,\mu) = R_{\mu \nu \lambda \sigma} R^{\mu \nu \lambda \sigma} = \left\{
\begin{array}{ll}
1/r_H^4 & \mu r_H \ll 1 \cr
1/\mu^2 r_H^6 & \mu r_H \gg  1 \end{array} \right. \  . \ee
One then confirms that the dimensionless quantity
\be R^2(r_H, \mu) T_H(\mu)^4 \ , \ee
at expected cross-over radius $r_H ={1 / T_H(\mu)}$, is always of
order one regardless of the value of $\mu l_s$ as long as one also
treats $\log(\mu l_s)$ as a quantity of order one. It appears
therefore that the correspondence principle is indeed working even
though both the Hagedorn density and the curvature depend
non-trivially on $\mu$.

One additional condition, which was assumed implicitly in this
discussion, requires that the string coupling be sufficiently weak, in
order to ensure that the entropy of excited strings at the cross-over
scale
\be {E \over T_H}  = {r_H^6 \over G_9 T_H} = {1 \over g^2 l_s^7 T_H^7}  \gg 1\ee%
is macroscopic.

\section{Discussion}

In this article, we computed the entropy as a function of energy for
black strings in an asymptotically plane wave background and for
string theory in the same plane-wave background.  In fact, the thermal
partition function can be computed for any of the ``exactly solvable''
backgrounds (\ref{background}) in string theory considered in
\cite{Russo:1995tj,Russo:1996ik}, by substituting appropriate values
to the general expression for the thermal partition function derived
in section 4.  For example, plane wave metric written in the form
\be ds^2 = -d  t^2 + d  y^2 + \sum_i (d  \rho_i^2 + 2 \eta \,  \rho_i^2 \,  d  \phi_i (d  t + d  y)) + 
\sum_{i=1}^4  (d \rho_i^2 + \rho_i^2 d \phi_i^2) \ee
which is useful for relating plane waves to G\"odel universes, 
corresponds to setting
\be \alpha_i = 2 \eta, \qquad \beta_i = 0, \qquad  q_{+i} = 0,\qquad q_{-i} = 0 \ . \ee
Other related backgrounds, such as the Melvin universe, can also be
considered.  Indeed, a large class of black string solution in various
asymptotic geometries can be constructed using a generalization of the
Null Melvin Twist, and for each of these solutions, the corresponding
thermal partition function for the free strings can be computed and
compared.

A rigorous definition of mass and energy in a non-asymptotically flat
space-time, however, can be rather subtle.  In this paper, we used the
notion of energy implied by the thermodynamic relation $dE = T dS$,
and found that the correspondence principle is working well with this
definition.  This however should not be considered as an acceptable
substitute for a careful definition of mass and energy in these
spaces, and we hope that a more satisfying formulation of these
quantities would appear in the literature in a due course (and
hopefully in agreement with our expectations).

\section*{Acknowledgments}

We would like to thank
E.~Gimon,
V.~Hubeny,
O.~Lunin,
M.~Rangamani,
J.~Russo,
G.~Shiu,
Y.~Sugawara,
A.~Tseytlin,
and D.~Vaman
for discussions. The work of AH was supported in part by funds from
the University of Wisconsin. The work of LAPZ was supported in part by
the DOE grant DE-FG02-95ER40899. AH thanks the EFI at the University
of Chicago and the MCTP at the University of Michigan, and LAPZ thanks
the University of Wisconsin-Madison, for hospitality while part of
this work was done.

\section*{Appendix A: Detailed Computation of the Thermal Partition Function}
         {\setcounter{section}{1} \gdef\thesection{\Alph{section}}}
          {\setcounter{equation}{0}}

In this appendix, we describe the computation of thermal partition
function. We follow much of the notations of \cite{Russo:1995tj} to
which readers are referred for additional information.

Thermal partition function can be computed in the path integral
formalism with only a minor modifications of the vacuum 1-loop
partition function.  Let us consider the case where only one set of
$\alpha$, $\beta$, $q_+$, and $q_-$ are non-zero for the sake of
illustration.  One loop partition function was computed in
\cite{Russo:1995tj} and is given by
\beq Z_0 &=& {r V_0 V_{22} \over (2 \pi)^{23}\alpha'^{23/2}}  \int_{{\cal F}} {d^2 \tau \over \tau^{2}_{2}}   \, 
  \sum_{w,w'=-\infty}^\infty  \left( 4 \tau_2^{-1} \int d \lambda d \bar \lambda \right)  
 \left( {e^{4 \pi \tau_2} \over \tau_2^{12}}    |f(e^{2 \pi i \tau})|^{-48} \right) \cr
&& {1 \over (2 \pi)^2} \exp[-{\pi(\chi - \tilde \chi)^2 \over 2  \tau_2} ] {{\tau_2} |\theta'_1(0,\tau)|^2 \over \theta_1(\chi| \tau) \theta_1(\tilde \chi| \bar \tau)}  \\
&& \exp\left( - 4 \pi \tau_2^{-1} \left(\lambda \bar \lambda 
- {1\over 2} r (w' - \tau w) \bar \lambda 
+ {1\over 2} r (w' - \bar \tau w) \lambda
\right) \right) \nonumber  \ . 
\eeq
where $\chi$ is
\beq 
\chi & = & - \sqrt{\alpha'} [2 \beta \lambda  + q_+ r(w' - \tau w)]\cr
\tilde \chi & = &  -\sqrt{\alpha'} [2 \alpha  \bar \lambda  + q_+ r(w' - \bar \tau w)] \ , 
\eeq
$r = R/\sqrt{\alpha'}$ is the radius of the compact $y$ direction in
string units, and $V_0$ is the infinite volume factor associated with
the time direction. To compute the thermal partition function,
evaluate the path integral around the background with boundary
condition
\be t(\sigma_1+n, \sigma_2+m) = t(\sigma_1,\sigma_2) + {1 \over T} m \ee
and integrate the modular parameter over the half-strip ${\cal E}$
instead of the fundamental domain ${\cal F}$. This makes the partition
function take the form
\beq Z_T &=& {r V_{22} \over (2 \pi)^{23}\alpha'^{23/2} T}  \int_{{\cal E}} {d^2 \tau \over \tau^{2}_{2}}   \, 
  \sum_{w,w'=-\infty}^\infty  \left( 4 \tau_2^{-1} \int d \lambda d \bar \lambda \right)  
 \left( {e^{4 \pi \tau_2} \over \tau_2^{12}}    |f(e^{2 \pi i \tau})|^{-48} \right) \cr
&& {1 \over (2 \pi)^2} \exp[-{\pi(\chi - \tilde \chi)^2 \over 2  \tau_2} ] {{\tau_2} |\theta'_1(0,\tau)|^2 \over \theta_1(\chi| \tau) \theta_1(\tilde \chi| \bar \tau)}  \label{final1a}\\
&& \exp\left( - 4 \pi \tau_2^{-1} \left(\lambda \bar \lambda 
- \left({1\over 2} r (w' - \tau w)-{i \over 4 \pi \sqrt{\alpha'} T} \right) \bar \lambda  \right. \right. \cr
&& \left. \left.  \qquad \qquad 
+ \left({1\over 2} r (w' - \bar \tau w) + {i \over 4 \pi \sqrt{\alpha'} T}\right)\lambda
\right) \right) \nonumber  \ . 
\eeq
where now
\beq 
\chi & = & - \sqrt{\alpha'} [2 \beta \lambda  + q_+ r(w' - \tau w)] + {i q_- \over 2 \pi T} \cr
\tilde \chi & = &  -\sqrt{\alpha'} [2 \alpha  \bar \lambda  + q_+ r(w' - \bar \tau w)]  + {i q_- \over 2 \pi T} \ .  \label{chirulea}
\eeq
Using the trick of \cite{O'Brien:1987pn}, this partition function can be recast in a manifestly modular invariant form
\beq Z_T &=& {r V_{22} \over (2 \pi)^{23}\alpha'^{23/2} T}  \int_{{\cal F}} {d^2 \tau \over \tau^{2}_{2}}   \, 
  \sum_{w,w'=-\infty}^\infty  \left( 4 \tau_2^{-1} \int d \lambda d \bar \lambda \right)  
 \left( {e^{4 \pi \tau_2} \over \tau_2^{12}}    |f(e^{2 \pi i \tau})|^{-48} \right) \cr
&& {1 \over (2 \pi)^2} \exp[-{\pi(\chi - \tilde \chi)^2 \over 2  \tau_2} ] {{\tau_2} |\theta'_1(0,\tau)|^2 \over \theta_1(\chi| \tau) \theta_1(\tilde \chi| \bar \tau)}  \cr
&& \exp\left( - 4 \pi \tau_2^{-1} \left(\lambda \bar \lambda 
- \left({1\over 2} r (w' - \tau w)+{i (k' - \tau k) \over 4 \pi \sqrt{\alpha'} T} \right) \bar \lambda  \right. \right. \\
&& \left. \left. \qquad
+ \left({1\over 2} r (w' - \bar \tau w) - {i(k' - \bar \tau k) \over 4 \pi \sqrt{\alpha'} T}\right)\lambda
\right) \right) \nonumber    \\
\chi & = & - \sqrt{\alpha'} [2 \beta \lambda  + q_+ r(w' - \tau w)] + {i q_- (k' - \tau k)\over 2 \pi T} \cr
\tilde \chi & = &  -\sqrt{\alpha'} [2 \alpha  \bar \lambda  + q_+ r(w' - \bar \tau w)]  + {i q_- (k' - \bar \tau k) \over 2 \pi T} \ .  \nonumber 
\eeq
This expression is modular invariant. To see this, act with transformation
\be \tau \rightarrow -{1 \over \tau}, \quad 
\lambda \rightarrow {\lambda \over \tau}, \qquad 
\bar \lambda \rightarrow{\bar \lambda \over \bar \tau}\ , \label{modulartrans1}\ee 
\be (k,k')
\rightarrow (k',-k), \quad (w,w') \rightarrow (w',-w) \ . \label{modulartrans2}\ee
which also causes $\chi$  and $\tilde \chi$ to transform according to
\be \chi \rightarrow {\chi \over \tau},  \quad
\tilde \chi \rightarrow {\tilde \chi \over \bar \tau}\ . \label{chitrans} \ee
Of course, everything other than the $\tau$ is just an integration
variable, so this implies that the integrand of the $\tau$ integral,
when all of the other integrals are done, is a modular invariant
function of $\tau$.
In the most general case where $\alpha_i$, $\beta_i$, $q_{+i}$, and
$q_{-i}$ for all 12 planes are non-vanishing, one finds
\beq Z_T &=& {r \over (2 \pi)\alpha'^{1/2} T}  \int_{{\cal F}} {d^2 \tau \over \tau^{2}_{2}}   \, 
  \sum_{w,w'=-\infty}^\infty  \left( 4 \tau_2^{-1} \int d \lambda d \bar \lambda \right)  
 \left({e^{4 \pi \tau_2} \over \tau_2^{12}}  |f(e^{2 \pi i \tau})|^{-48} \right) \cr
&& \prod_{i=1}^{12} \left({1 \over (2 \pi)^2} \exp[-{\pi(\chi_i - \tilde \chi_i)^2 \over 2  \tau_2} ] {{\tau_2} |\theta'_1(0,\tau)|^2 \over \theta_1(\chi_i| \tau) \theta_1(\tilde \chi_i| \bar \tau)}  \right) \cr
&& \exp\left( - 4 \pi \tau_2^{-1} \left(\lambda \bar \lambda 
- \left({1\over 2} r (w' - \tau w)+{i (k' - \tau k) \over 4 \pi \sqrt{\alpha'} T} \right) \bar \lambda  \right. \right. \\
&& \left. \left. \qquad
+ \left({1\over 2} r (w' - \bar \tau w) - {i(k' - \bar \tau k) \over 4 \pi \sqrt{\alpha'} T}\right)\lambda
\right) \right) \nonumber    \\
\chi_i & = & - \sqrt{\alpha'} [2 \beta_i \lambda  + q_{+i} r(w' - \tau w)] + {i q_{-i} (k' - \tau k)\over 2 \pi T} \cr
\tilde \chi_i & = &  -\sqrt{\alpha'} [2 \alpha_i  \bar \lambda  + q_{+i} r(w' - \bar \tau w)]  + {i q_{-i} (k' - \bar \tau k) \over 2 \pi T} \ .  \nonumber 
\eeq
As a check, note that in the limit where all of the $\alpha_i$,
$\beta_i$, $q_{+i}$, and $q_{-i}$ goes to zero, there will be a
diverging factor of ${1 / (2 \pi)^2 (\chi_i \tilde \chi_i)}$ for each
of the 12 transverse planes coming from the contribution of the
zero-mode to the path integral. Replacing this factor with the
standard factor of ${V_2 / (2 \pi)^2 \alpha' \tau_2}$, we recover the
thermal partition function of bosonic strings in the conventional
normalization.

Thermal partition function for similar exactly solvable type II
backgrounds can also be computed along these lines.

\section*{Appendix B: Oscillator Computation of the Thermal Partition Function}
         {\setcounter{section}{2} \gdef\thesection{\Alph{section}}}
          {\setcounter{equation}{0}}

In this appendix, we describe the computation of the thermal partition
function using the oscillator approach. The goal is to explicitly
evaluate
\be Z_T = \left. {1 \over T} \int_{{\cal E}} {d^2 \tau \over \tau_2} \sum_{m,w} \int {d \varepsilon \over 2 \pi} e^{i k' \varepsilon / T} {\rm Tr} \left( e^{2 \pi i (\tau L_0 - \bar \tau \bar L_0)} \right) \right|_{k'=-1} \ee
for the background (\ref{background}).  Similar computation for the
vacuum partition function was originally done by \cite{Russo:1995tj}.
We therefore refer the reader to \cite{Russo:1995tj} for more
information regarding conventions and notations.  In actually
evaluating the partition function, we first trace over the oscillators
and then integrate over the zero modes. This is opposite of the order
in which the computation was done in \cite{Russo:1995tj}. Doing the
trace first actually clarifies certain technical aspect of this
computation.

Let describe the case where only one set of $\alpha_i$, $\beta_i$,
$q_{+i}$, and $q_{-i}$ is non-vanishing only for $i=1$. One must then
include the zero mode integral over $p^a$'s
\be Z_T = \left. {V_{22} \over T} \int_{{\cal E}} {d^2 \tau \over \tau_2} \sum_{m,w} \int {d \varepsilon \over 2 \pi} {d^{22} p \over (2 \pi)^{22}} e^{i k' \varepsilon / T} {\rm Tr} \left( e^{2 \pi i (\tau L_0 - \bar \tau \bar L_0)} \right) \right|_{k'=-1} \ . \label{definition} \ee
The $\hat L_0$ and the $\hat {\bar L}_0$ operators are given by
\beq \hat L_0
& = & {p_-^u p_-^v \over 4 \alpha'}  + {\alpha' p_a^2 \over 4} + N - {1 \over 2}\gamma' (\hat J_R + {1 \over 2}) - c_0 \cr
\hat {\bar L}_0
& = & {p_+^u p_+^v \over 4 \alpha'}  + {\alpha' p_a^2 \over 4}+ \bar N + {1 \over 2}\gamma' (\hat J_L - {1 \over 2}) - c_0 \ . \eeq
where 
\beq  c_0 &=& 1 - {1 \over 4} \gamma' + {1 \over 8} \gamma'^2 \cr
\hat N & = & \sum_{n=1}^\infty n(
b^\dag_{n+} b_{n+} +  b^\dag_{n-} b_{n-} + a^\dag_{na} a_{na}) \cr
\hat {\bar N} & = & \sum_{n=1}^\infty n(
\tilde b^\dag_{n+} \tilde b_{n+} +  \tilde b^\dag_{n-} \tilde b_{n-} + \tilde a^\dag_{na} \tilde a_{na}) \cr
\hat J_R & = &  -b_0^\dag b_0 - {1 \over 2} + \sum_{n=1}^\infty (b_{n+}^\dag b_{n+}  - b_{n-}^\dag b_{n-} )\cr
\hat J_L & = &  \tilde b_0^\dag \tilde b_0 + {1 \over 2} + \sum_{n=1}^\infty (\tilde b_{n+}^\dag \tilde b_{n+}  - \tilde b_{n-}^\dag \tilde b_{n-} )\cr
{p^u_- p^v_- \over 4 \alpha'}+ {p^u_+ p^v_+ \over 4 \alpha'} 
&=& {\alpha' \over 2} \left( - \left(E - {(a_-+c_-) \hat J \over 2} \right)^2  
 + \left(p_y  - {(a_+ + c_+) \hat J \over 2} \right)^2 \right. \cr
&& \left . + \left( {w R \over \alpha'} - { (a_+ + c_+) \hat J \over 2} \right)^2 -  {(a_- - c_-)^2 \hat J^2 \over 4}  \right) \cr
{p^u_- p^v_- \over 4 \alpha'}- {p^u_+ p^v_+ \over 4 \alpha'}  & = &
- m w + {\gamma \hat J  \over 2}   \ , \eeq
and
\beq 
\gamma &=& (c_+ + a_+) w R + {1 \over 2} (\alpha + \beta) s + {1 \over 2 } (\alpha - \beta) p\cr
\gamma' & = & 2({1 \over 2} \gamma  - [{1 \over 2} \gamma])  \ . 
\eeq
The fact that the $\hat L_0$ operators depend on $\gamma'$ which not a
continuous function of $\gamma$ may seem to suggest that the
evaluation of the partition function cumbersome. One can however show
that this expression is invariant under shift of $\gamma'$ by 2, by
computing the trace over oscillators first. To address this issue, let
us introduce auxiliary variables ${\cal J}_L$ and ${\cal J}_R$ and
write
\beq Z_T &=& {V_{22} \over T}
\int d {\cal J}_L \int d {\cal J}_R \int {d^2 \tau \over \tau_{2}} \int_{-\infty}^{\infty} {d \varepsilon \over 2 \pi}\,  {d^{22} p \over (2 \pi)^{22}} \sum_{m,w=-\infty}^\infty \cr
&& \left. e^{{i k' \varepsilon / T}} {\rm Tr}  \delta(\hat J_L - {\cal J}_L) \delta(\hat J_R - {\cal J}_R)\exp[ 2 \pi i (\tau L_0 - \bar\tau \bar L_0)] \rule{0ex}{3ex}\right|_{k'=-1} \ , \eeq
introduce an integral expression for the delta function
\be \delta(\hat J_R - {\cal J}_R) = \int d \chi \exp[- 2 \pi i \chi (J_R - {\cal J}_R)] \ , \qquad
\delta(\hat J_L - {\cal J}_L) = \int d \tilde \chi \exp[- 2 \pi i \tilde \chi (J_L - {\cal J}_L)] \ , \ee
and write $\gamma$ in terms of ${\cal J}_{L,R}$ as
\be
\gamma = (a_+ + c_+) w R + \alpha' [(c_+ - a_+) p_y + (a_- - c_-)E] + {1 \over 2} \alpha' (a_+^2 - a_-^2 - c_+^2 + c_-^2) ({\cal J}_L + {\cal J}_R) 
\ . \ee
Using the trick of writing
\be \exp\left({\pi \tau_2 \over 2}  \gamma^2 \right) =  \sqrt{{\tau_2 \over 2}}\int d\nu \, \exp\left( - {1 \over 2} \pi \tau^2 \nu^2 - \pi \tau^2 \nu \gamma' \right)\,  \ee
introducing shift of variables
\be {\cal J_L} = {\cal J}'_L - {1 \over 2 } \nu, \qquad 
{\cal J_R} = {\cal J}'_R + {1 \over 2 } \nu \ , \ee
and integrating out $\nu$, one can show that
\beq 
Z_T &=&  {V_{22} \over T} \int {d\tau^2 \over \tau_{2}} \int d\chi  \int d \tilde \chi   \int d {\cal J}'_L \int d {\cal J}'_R  \int_{-\infty}^{\infty} {d \varepsilon \over 2 \pi} \, {d^a p \over (2 \pi)^a} \sum_{m,w=-\infty}^\infty e^{ {i k'  \varepsilon / T}}   \exp \left( - {\pi (\chi - \tilde \chi)^2 \over 2  \tau_2}  \right)  \cr
&&
\left( {\rm Tr} \exp\left[ 2 \pi i (\tau (N-1) -  \chi \hat J_R)\right]\right)
\left( {\rm Tr} \exp\left[ -2 \pi i (\bar \tau (\bar N - 1) + \tilde \chi \hat J_L)\right]\right)  \label{eq} \\
&& 
\exp\left[\rule{0ex}{3ex} 2 \pi i \chi {\cal J}'_R + 2 \pi i \tilde \chi {\cal J}'_L
 - \pi i \tau \gamma' {\cal J}'_R  
 - \pi i \bar \tau \gamma' {\cal J}'_L 
\right. \cr
&& \left. \left.\rule{0ex}{3ex}
\qquad + 2 \pi i \tau \left({p_-^u p_-^v \over 4 \alpha'}   + {\alpha' p_a^2 \over 4} \right)
- 2 \pi i \bar\tau \left({p_+^u p_+^v \over 4 \alpha'}  + {\alpha' p_a^2 \over 4} \right) \right]  \right|_{k'=-1} \ . \nonumber
\eeq
The trace can be computed
\beq 
\lefteqn{{\rm Tr} \left\{ \rule{0ex}{4ex}
 \exp\left[
2 \pi i (\tau (N-1) - \chi \hat J'_R)
-2 \pi i (\bar \tau (\bar N  -1) + \tilde \chi \hat J'_L)
\right]  \right\} } \cr
&  =& {1 \over (2 \pi)^2} e^{4 \pi \tau_2} |f(e^{2 \pi i \tau})|^{-48}  e^{i \pi(\chi - \tilde \chi) \nu} {|\theta'_1(0,\tau)|^2 \over \theta_1(\chi,\tau) \theta_1(\tilde \chi, \bar\tau)}  \ , 
\eeq
giving
\beq 
Z_T &=&  {V_{22} \over T} \int {d\tau^2 \over \tau_{2}} \int d\chi  \int d \tilde \chi   \int d {\cal J}'_L \int d {\cal J}'_R  \int_{-\infty}^{\infty} {d \varepsilon \over 2 \pi} \, {d^a p \over (2 \pi)^a} \sum_{m,w=-\infty}^\infty e^{ {i k' \varepsilon / t}}   \cr
&&
\left( {1 \over (2 \pi)^2} e^{4 \pi \tau_2} |f(e^{2 \pi i \tau})|^{-48}  \exp \left( - {\pi (\chi - \tilde \chi)^2 \over 2  \tau_2}  \right)   {|\theta'_1(0,\tau)|^2 \over \theta_1(\chi,\tau) \theta_1(\tilde \chi, \bar\tau)}  \right) \\
&& 
\exp\left[\rule{0ex}{3ex} 2 \pi i \chi {\cal J}'_R + 2 \pi i \tilde \chi {\cal J}'_L
 - \pi i \tau \gamma' {\cal J}'_R  
 - \pi i \bar \tau \gamma' {\cal J}'_L 
\right. \cr
&& \left. \left.\rule{0ex}{3ex}
+ 2 \pi i \tau \left({p_-^u p_-^v \over 4 \alpha'} + {\alpha' p_a^2 \over 4}    \right)
- 2 \pi i \bar\tau \left({p_+^u p_+^v \over 4 \alpha'}  + {\alpha' p_a^2 \over 4} \right) \right]  \right|_{k'=-1} \ . \nonumber
\eeq
In this form, it can be readily verified that a shift $\gamma'
\rightarrow \gamma' + 2$ can be canceled by a shift of integration
variables
\be \chi \rightarrow \chi +  \tau, \qquad \tilde \chi \rightarrow \bar \chi + \bar \tau \ .  \ee
Now that we see that $\gamma'$ can freely be replaced by $\gamma$, let
us integrate out the zero modes $\varepsilon$ and $p_a$, which is a
Gaussian integral, and Poisson resum $m$. To evaluate the integrals
further, it is convenient to introduce an auxiliary variable by
multiplying the integrand by
\beq \lefteqn{ 1 = 4 \tau_2^{-1} \int d \lambda d \bar \lambda} \label{auxiliary} \\
&& \qquad \exp\left( - 4 \pi \tau_2^{-1} [\lambda - {1 \over 2} r(w' - \tau w) + i \tau_2 \sqrt{\alpha'} \alpha {\cal J}'_L] [\bar \lambda + {1 \over 2} r(w'-\bar \tau w) - i \tau_2 \sqrt{\alpha'} \beta {\cal J}'_R] \right) \ .\nonumber \eeq
Then, ${\cal J}'_L$ and ${\cal J}'_R$ integrals give rise to a delta
function which constrain $\chi$ and $\tilde \chi$. As a result of
these integrations, one finally arrives at
\beq Z_T &=& {r V_{22} \over (2 \pi)^{23}\alpha'^{23/2} T}  \int_{{\cal E}} {d^2 \tau \over \tau^{2}_{2}}   \, 
  \sum_{w,w'=-\infty}^\infty  \left( 4 \tau_2^{-1} \int d \lambda d \bar \lambda \right)  
 \left( {e^{4 \pi \tau_2} \over \tau_2^{12}}  |f(e^{2 \pi i \tau})|^{-48} \right) \cr
&&  {1 \over (2 \pi)^2} \exp[-{\pi(\chi - \tilde \chi)^2 \over 2 \tau_2} ]  {\tau_2 |\theta'_1(0,\tau)|^2 \over \theta_1(\chi| \tau) \theta_1(\tilde \chi| \bar \tau)}  \label{final}\\
&& \left. \exp\left( - 4 \pi \tau_2^{-1} \left({k'^2 \over 16 \pi^2 \alpha' T^2} + \lambda \bar \lambda 
- {1\over 2} r (w' - \tau w) \bar \lambda 
+ {1\over 2} r (w' - \bar \tau w) \lambda
\right) \right) \right|_{k'=-1} \nonumber  \ . 
\eeq
where
\beq 
\chi & = & - \sqrt{\alpha'} [2 \beta \lambda  + q_+ r(w' - \tau w)] + {i c_- k' \over 2 \pi T} \cr
\tilde \chi & = &  -\sqrt{\alpha'} [2 \alpha  \bar \lambda  + q_+ r(w' - \bar \tau w)] + {i a_- k' \over 2 \pi T} \ .
\eeq
Up to a shift in $\lambda$ and $\bar \lambda$, this is identical to
(\ref{final1a}) and (\ref{chirulea}) that was presented in the
previous section.

\bibliography{hagedorn}\bibliographystyle{utphys}

\providecommand{\href}[2]{#2}\begingroup\raggedright\begin{thebibliography}{10}

\bibitem{Russo:1995cv}
J.~G. Russo and A.~A. Tseytlin, ``Constant magnetic field in closed string
  theory: an exactly solvable model,'' {\em Nucl. Phys.} {\bf B448} (1995)
  293--330,
\href{http://www.arXiv.org/abs/hep-th/9411099}{{\tt hep-th/9411099}}.

\bibitem{Metsaev:2001bj}
R.~R. Metsaev, ``Type IIB Green-Schwarz superstring in plane wave Ramond-Ramond
  background,'' {\em Nucl. Phys.} {\bf B625} (2002) 70--96,
\href{http://www.arXiv.org/abs/hep-th/0112044}{{\tt hep-th/0112044}}.

\bibitem{Blau:2001ne}
M.~Blau, J.~Figueroa-O'Farrill, C.~Hull, and G.~Papadopoulos, ``A new maximally
  supersymmetric background of IIB superstring theory,'' {\em JHEP} {\bf 01}
  (2002) 047,
\href{http://www.arXiv.org/abs/hep-th/0110242}{{\tt hep-th/0110242}}.

\bibitem{Berenstein:2002jq}
D.~Berenstein, J.~M. Maldacena, and H.~Nastase, ``Strings in flat space and
  pp-waves from ${\cal N}$ = 4 super Yang Mills,'' {\em JHEP} {\bf 04} (2002)
  013,
\href{http://www.arXiv.org/abs/hep-th/0202021}{{\tt hep-th/0202021}}.

\bibitem{Witten:1998zw}
E.~Witten, ``Anti-de Sitter space, thermal phase transition, and confinement in
  gauge theories,'' {\em Adv. Theor. Math. Phys.} {\bf 2} (1998) 505--532,
\href{http://www.arXiv.org/abs/hep-th/9803131}{{\tt hep-th/9803131}}.

\bibitem{Hubeny:2003ug}
V.~E. Hubeny and M.~Rangamani, ``Horizons and plane waves: a review,'' {\em
  Mod. Phys. Lett.} {\bf A18} (2003) 2699--2712,
\href{http://www.arXiv.org/abs/hep-th/0311053}{{\tt hep-th/0311053}}.

\bibitem{Gimon:2003ms}
E.~G. Gimon and A.~Hashimoto, ``Black holes in G{\oo}del universes and
  pp-waves,'' {\em Phys. Rev. Lett.} {\bf 91} (2003) 021601,
\href{http://www.arXiv.org/abs/hep-th/0304181}{{\tt hep-th/0304181}}.

\bibitem{Gimon:2003xk}
E.~G. Gimon, A.~Hashimoto, V.~E. Hubeny, O.~Lunin, and M.~Rangamani, ``Black
  strings in asymptotically plane wave geometries,'' {\em JHEP} {\bf 08} (2003)
  035,
\href{http://www.arXiv.org/abs/hep-th/0306131}{{\tt hep-th/0306131}}.

\bibitem{Herdeiro:2002ft}
C.~A.~R. Herdeiro, ``Spinning deformations of the D1-D5 system and a geometric
  resolution of closed timelike curves,'' {\em Nucl. Phys.} {\bf B665} (2003)
  189--210,
\href{http://www.arXiv.org/abs/hep-th/0212002}{{\tt hep-th/0212002}}.

\bibitem{Bena:2002kq}
I.~Bena and R.~Roiban, ``Supergravity pp-wave solutions with 28 and 24
  supercharges,'' {\em Phys. Rev.} {\bf D67} (2003) 125014,
\href{http://www.arXiv.org/abs/hep-th/0206195}{{\tt hep-th/0206195}}.

\bibitem{Michelson:2002ps}
J.~Michelson, ``A pp-wave with 26 supercharges,'' {\em Class. Quant. Grav.}
  {\bf 19} (2002) 5935--5949,
\href{http://www.arXiv.org/abs/hep-th/0206204}{{\tt hep-th/0206204}}.

\bibitem{Susskind:1993ws}
L.~Susskind, ``Some speculations about black hole entropy in string theory,''
\href{http://www.arXiv.org/abs/hep-th/9309145}{{\tt hep-th/9309145}}.

\bibitem{Horowitz:1997nw}
G.~T. Horowitz and J.~Polchinski, ``A correspondence principle for black holes
  and strings,'' {\em Phys. Rev.} {\bf D55} (1997) 6189--6197,
\href{http://www.arXiv.org/abs/hep-th/9612146}{{\tt hep-th/9612146}}.

\bibitem{Li:2002mq}
M.~Li, ``Correspondence principle in a pp-wave background,'' {\em Nucl. Phys.}
  {\bf B638} (2002) 155--164,
\href{http://www.arXiv.org/abs/hep-th/0205043}{{\tt hep-th/0205043}}.

\bibitem{Alishahiha:2003ru}
M.~Alishahiha and O.~J. Ganor, ``Twisted backgrounds, pp-waves and nonlocal
  field theories,'' {\em JHEP} {\bf 03} (2003) 006,
\href{http://www.arXiv.org/abs/hep-th/0301080}{{\tt hep-th/0301080}}.

\bibitem{Russo:1995tj}
J.~G. Russo and A.~A. Tseytlin, ``Exactly solvable string models of curved
  space-time backgrounds,'' {\em Nucl. Phys.} {\bf B449} (1995) 91--145,
\href{http://www.arXiv.org/abs/hep-th/9502038}{{\tt hep-th/9502038}}.

\bibitem{Russo:1996ik}
J.~G. Russo and A.~A. Tseytlin, ``Magnetic flux tube models in superstring
  theory,'' {\em Nucl. Phys.} {\bf B461} (1996) 131--154,
\href{http://www.arXiv.org/abs/hep-th/9508068}{{\tt hep-th/9508068}}.

\bibitem{Brace:2003st}
D.~Brace, C.~A.~R. Herdeiro, and S.~Hirano, ``Classical and quantum strings in
  compactified pp-waves and G{\oo}del type universes,''
\href{http://www.arXiv.org/abs/hep-th/0307265}{{\tt hep-th/0307265}}.

\bibitem{Wald:1984rg}
R.~M. Wald, {\em General Relativity}.
\newblock Chicago University Press, 1984.

\bibitem{Takayanagi:2001jj}
T.~Takayanagi and T.~Uesugi, ``Orbifolds as Melvin geometry,'' {\em JHEP} {\bf
  12} (2001) 004,
\href{http://www.arXiv.org/abs/hep-th/0110099}{{\tt hep-th/0110099}}.

\bibitem{Dudas:2001ux}
E.~Dudas and J.~Mourad, ``D-branes in string theory Melvin backgrounds,'' {\em
  Nucl. Phys.} {\bf B622} (2002) 46--72,
\href{http://www.arXiv.org/abs/hep-th/0110186}{{\tt hep-th/0110186}}.

\bibitem{Takayanagi:2001aj}
T.~Takayanagi and T.~Uesugi, ``D-branes in Melvin background,'' {\em JHEP} {\bf
  11} (2001) 036,
\href{http://www.arXiv.org/abs/hep-th/0110200}{{\tt hep-th/0110200}}.

\bibitem{Takayanagi:2002je}
H.~Takayanagi and T.~Takayanagi, ``Open strings in exactly solvable model of
  curved space-time and pp-wave limit,'' {\em JHEP} {\bf 05} (2002) 012,
\href{http://www.arXiv.org/abs/hep-th/0204234}{{\tt hep-th/0204234}}.

\bibitem{Takayanagi:2002pi}
T.~Takayanagi, ``Modular invariance of strings on pp-waves with RR-flux,'' {\em
  JHEP} {\bf 12} (2002) 022,
\href{http://www.arXiv.org/abs/hep-th/0206010}{{\tt hep-th/0206010}}.

\bibitem{PandoZayas:2002hh}
L.~A. Pando~Zayas and D.~Vaman, ``Strings in RR plane wave background at finite
  temperature,'' {\em Phys. Rev.} {\bf D67} (2003) 106006,
\href{http://www.arXiv.org/abs/hep-th/0208066}{{\tt hep-th/0208066}}.

\bibitem{Greene:2002cd}
B.~R. Greene, K.~Schalm, and G.~Shiu, ``On the Hagedorn behaviour of pp-wave
  strings and ${\cal N}$ = 4 SYM theory at finite R-charge density,'' {\em
  Nucl. Phys.} {\bf B652} (2003) 105--126,
\href{http://www.arXiv.org/abs/hep-th/0208163}{{\tt hep-th/0208163}}.

\bibitem{Sugawara:2002rs}
Y.~Sugawara, ``Thermal amplitudes in DLCQ superstrings on pp-waves,'' {\em
  Nucl. Phys.} {\bf B650} (2003) 75--113,
\href{http://www.arXiv.org/abs/hep-th/0209145}{{\tt hep-th/0209145}}.

\bibitem{Brower:2002zx}
R.~C. Brower, D.~A. Lowe, and C.-I. Tan, ``Hagedorn transition for strings on
  pp-waves and tori with chemical potentials,'' {\em Nucl. Phys.} {\bf B652}
  (2003) 127--141,
\href{http://www.arXiv.org/abs/hep-th/0211201}{{\tt hep-th/0211201}}.

\bibitem{Sugawara:2003qc}
Y.~Sugawara, ``Thermal partition function of superstring on compactified
  pp-wave,'' {\em Nucl. Phys.} {\bf B661} (2003) 191--208,
\href{http://www.arXiv.org/abs/hep-th/0301035}{{\tt hep-th/0301035}}.

\bibitem{Grignani:2003cs}
G.~Grignani, M.~Orselli, G.~W. Semenoff, and D.~Trancanelli, ``The superstring
  Hagedorn temperature in a pp-wave background,'' {\em JHEP} {\bf 06} (2003)
  006,
\href{http://www.arXiv.org/abs/hep-th/0301186}{{\tt hep-th/0301186}}.

\bibitem{Hyun:2003ks}
S.-J. Hyun, J.-D. Park, and S.-H. Yi, ``Thermodynamic behavior of IIA string
  theory on a pp-wave,'' {\em JHEP} {\bf 11} (2003) 006,
\href{http://www.arXiv.org/abs/hep-th/0304239}{{\tt hep-th/0304239}}.

\bibitem{Bigazzi:2003jk}
F.~Bigazzi and A.~L. Cotrone, ``On zero-point energy, stability and Hagedorn
  behavior of type IIB strings on pp-waves,'' {\em JHEP} {\bf 08} (2003) 052,
\href{http://www.arXiv.org/abs/hep-th/0306102}{{\tt hep-th/0306102}}.

\bibitem{Atick:1988si}
J.~J. Atick and E.~Witten, ``The Hagedorn transition and the number of degrees
  of freedom of string theory,'' {\em Nucl. Phys.} {\bf B310} (1988)
291--334.

\bibitem{O'Brien:1987pn}
K.~H. O'Brien and C.~I. Tan, ``Modular invariance of thermopartition function
  and global phase structure of heterotic string,'' {\em Phys. Rev.} {\bf D36}
  (1987)
1184.

\end{thebibliography}\endgroup

\end{document}